\documentclass[sigconf]{acmart}
\AtBeginDocument{%
  \providecommand\BibTeX{{%
    \normalfont B\kern-0.5em{\scshape i\kern-0.25em b}\kern-0.8em\TeX}}}

\copyrightyear{2021}
\acmYear{2021}
\setcopyright{iw3c2w3}
\acmConference[WWW '21 Companion]{Companion Proceedings of the Web Conference 2021}{April 19--23, 2021}{Ljubljana, Slovenia}
\acmBooktitle{Companion Proceedings of the Web Conference 2021 (WWW '21 Companion), April 19--23, 2021, Ljubljana, Slovenia}
\acmPrice{}
\acmDOI{10.1145/3442442.3451362}
\acmISBN{978-1-4503-8313-4/21/04}

\begin{document}

\title[On Representation Learning for Scientific News Articles]{On Representation Learning for Scientific News Articles \\ Using Heterogeneous Knowledge Graphs}


\author{Angelika Romanou}
\email{angelika.romanou@epfl.ch}
\affiliation{%
  \institution{
    École polytechnique fédérale de Lausanne (EPFL)}
  \country{Switzerland}
}

\author{Panayiotis Smeros}
\email{panayiotis.smeros@epfl.ch}
\affiliation{%
  \institution{
    École polytechnique fédérale de Lausanne (EPFL)}
  \country{Switzerland}
}

\author{Karl Aberer}
\email{karl.aberer@epfl.ch}
\affiliation{%
  \institution{
    École polytechnique fédérale de Lausanne (EPFL)}
  \country{Switzerland}
}

\renewcommand{\shortauthors}{Romanou et al.}

\begin{abstract}
In the era of misinformation and information inflation, the credibility assessment of the produced news is of the essence. 
However, fact-checking can be challenging considering the limited references presented in the news. 
This challenge can be transcended by utilizing the knowledge graph that is related to the news articles.
In this work, we present a methodology for creating scientific news article representations by modeling the directed graph between the scientific news articles and the cited scientific publications. 
The network used for the experiments is comprised of the scientific news articles, their topic, the cited research literature, and their corresponding authors. 
We implement and present three different approaches: 1) a baseline Relational Graph Convolutional Network (R-GCN), 2) a Heterogeneous Graph Neural Network (HetGNN) and 3) a Heterogeneous Graph Transformer (HGT). 
We test these models in the downstream task of link prediction on the: a) news article - paper links and b) news article - article topic links.
The results show promising applications of graph neural network approaches in the domains of knowledge tracing and scientific news credibility assessment. 

\end{abstract}



\maketitle

\section{Introduction}
\label{sec:intro}

In recent years, the news landscape has changed radically mainly due to the instantaneous rate at which media publish content as well as the lack of regulation and quality control.
This plethora of news sources has inevitably led to a burst of misinformation \cite{fernandez2018online}, increasing the need for more sophisticated approaches for mining and evaluating the news. 
Fact checking, viewpoints comparison, news categorization and news recommendation are some of the numerous use cases that computational journalism can be applied. 
Broad research has been done in the domain of computational journalism in order to provide a solid framework for news assessment \cite{smeros2019scilens}. 
An important aspect of analysing the news articles is the ability to build a mosaic of heterogeneous information that derive from various sources.

A Knowledge graph (KG) is a typical example of bringing together collections of interlinked descriptions of entities. 
KGs have witnessed rapid growth within the recent years with real-world applications around information extraction, question answering and entity disambiguation. 
A KG is a multi-relational graph composed of entities (nodes) and relations (different types of edges). 
Each edge in the knowledge graph is represented as a triple of the form \textit{(head entity, relation, tail entity)}, indicating that two entities are connected by a specific relation. 
Although effective in representing structured information, the complexity of KGs usually makes them hard to process. 
To tackle this issue, extensive research has been made to embed components of a KG including entities and relations into continuous vector spaces and creating graph embeddings, simplify the manipulation of KGs while preserving its structure. 
A detailed survey around the approaches and applications of KGs' embeddings is presented by \citet{wang2017knowledge}.

The task of meaningful vector representation for each node is challenging not only because of the need to consider heterogeneous structural graph information, but also due to the demand for incorporating heterogeneous attributes associated with each node. 
In the case of web knowledge graphs and more specifically news articles networks, the news article entities can have a very diverse type of related attributes such as the images of an article, the entities that are mentioned inside the corpus of the article, indicators and statistics about the credibility or the popularity of an article etc.
Furthermore, in real-world dynamic networks such as news graphs, the node or edge counts of different types can vary greatly through time. 
An inevitable effect of this temporal aspect when mining those networks, is the fact that for relations that might not have sufficient occurrences due to their recency, it's hard to learn accurate relation-specific representations. 
These challenges make the demand for versatile and effective mining of the web knowledge graphs more apparent.

In this work we aim to use heterogeneous graph embedding strategies in order to enhance the analysis toolbox around tackling news misinformation and fact assessment. 
We use a scientific news network as presented in \ref{fig:network}(a) which is comprised of the scientific news article entities along with their cited academic literature. 
We extract the representations for each news article and we showcase the effectiveness of this approach into two different use cases that are explained below. 

\subsection{Use Case A: Scientific News Article Evaluation}
Nowadays, science is communicated solely through information elicited from news articles and blog-posts. 
This is apparent to domains where scientific literature remains an exclusive privilege, locked behind paywalls.
With the burst of misinformation, the validity assessment of the popularized published science is of the essence. 
Unfortunately, a large amount of news articles that state scientific claims might not cite the original references from the scientific literature. 
This renders the non-expert audience unable to assess the credibility of the source or fact check its information. 
With the use of the scientific news network, we aim to connect the news article entities with potential related scientific literature based on their claims and their structural characteristics of the neighboring nodes in the network. 
In the first use case presented, we model the scientific news articles network and we work on predicting the edges connecting the news articles with their relevant scientific papers. 

\subsection{Use Case B: Scientific News Article Topic Classification}
A common use case regarding news article classification is the one of the article categorization into one or more topics that its content refers to.
This task is usually implemented by supervised approaches based on article implicit characteristics such as its title, its corpus or its incorporated images. 
In our case, we treat this problem as a link prediction problem in the scientific news graph, by incorporating the structural characteristics of the network.
Thus, in this second use case, we are predicting the edges between the news articles and the topic entities associated with it.

The rest of the document is structured in the following manner: Section \ref{sec:lr} presents the related work around heterogeneous graphs embedding methods; Section \ref{sec:network} describes the dataset used in this work; Section \ref{sec:methodology} explains the methodology for each implemented model; Section \ref{sec:results} presents the results of the models for each use case and finally, Section \ref{sec:conclusion} summarises the results and prompts for a discussion about future work and further applications.

\begin{figure}
  \centering
  \includegraphics[width=1\linewidth]{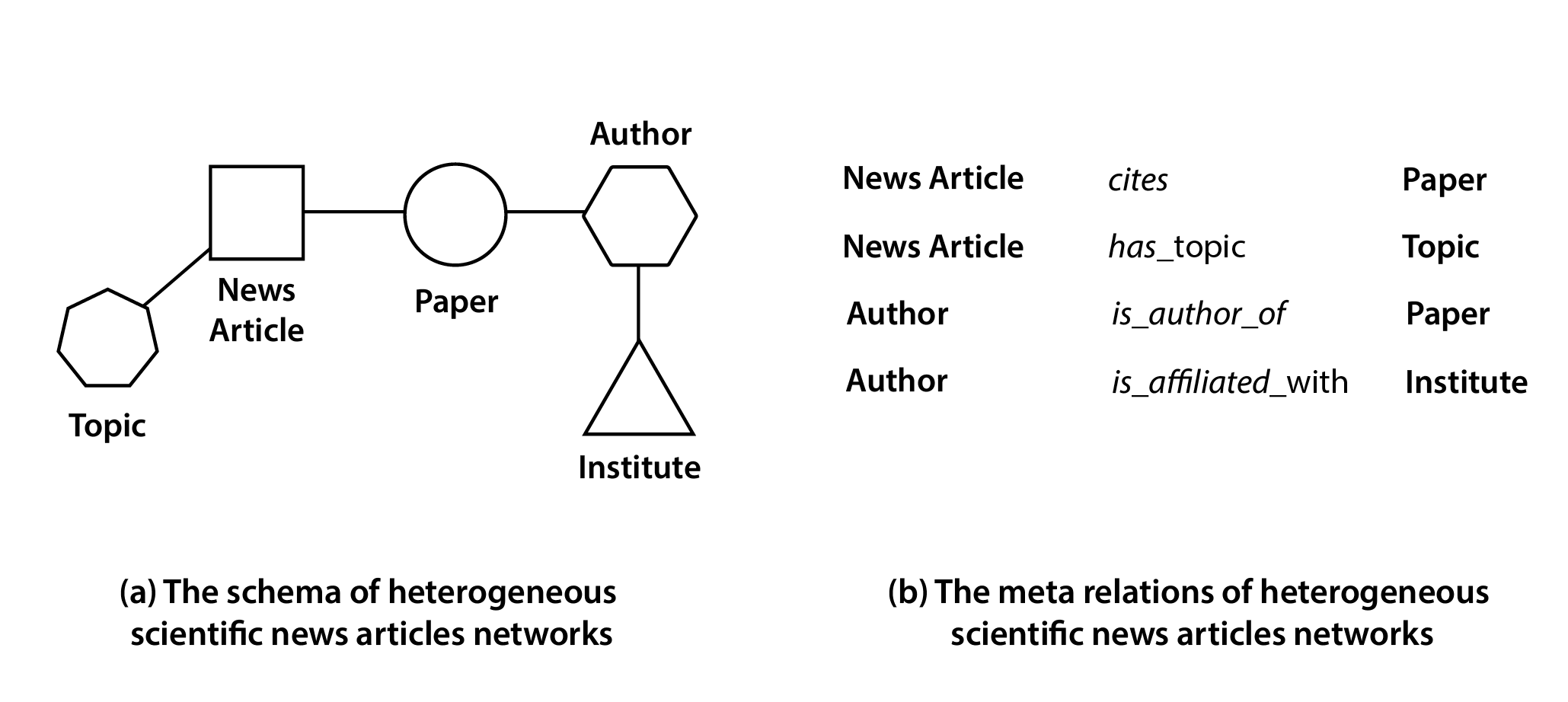}
  \caption{The schema and meta relations of scientific news articles graph.}
\label{fig:network}
\end{figure}

\section{Related Work}
\label{sec:lr}
In this section, we review the recent developments on heterogeneous graph representations and news articles embeddings 
and highlight the differences between our approach and the related literature. 

\subsection{Heterogeneous Graph Representation Learning}
Over the past decade, a great amount of work has been conducted for mining of heterogeneous graphs \cite{sun2012mining}. 
One of the most popular data mining topics in recent years has been the graph representation learning \cite{cui2018survey}.
One of the classical paradigms is to perform short random walks over the graph in order to learn latent node representations, such as Deepwalk \cite{perozzi2014deepwalk}.
Similarly, metapath2vec \cite{dong2017metapath2vec} designs a meta-path based random walk by utilizing skip-gram and HERec \cite{shi2018heterogeneous} captures the complex semantics by using a type constrain strategy to filter the node sequence. 
Additional graph structure based models were proposed to learn vectorized node embeddings that leverage both the graph structure and the node attributes \cite{li2017attributed, liao2018attributed} as well as methods that utilize matrix factorization \cite{qiu2018network} and adversarially regularized autoencoders \cite{zhao2018adversarially}. 
In this work, we extract heterogeneous graph representations to a web-scale use case of the scientific news article field.

\subsection{Heterogeneous Graph Neural Networks}
Recently, due to the success of the graph neural networks (GNN), several GNN variations has been published on modeling heterogeneous networks \cite{schlichtkrull2018modeling,wang2019heterogeneous,zhang2019heterogeneous,hu2020heterogeneous}. 
Compared to previous graph word embedding models, GNNs are able to aggregate information from node's local neighbors via neural networks leveraging the context of each entity type. 
For example, the Relational Graph Convolutional Networks (R-GCN) \cite{schlichtkrull2018modeling} apply GNN framework on relational data by having different projection matrices for each unique type of relation in the network. 
Later, an extension of this idea has been introduced with the HetGNNs \cite{zhang2019heterogeneous} which combine the contextual information of the graph with the node attributes for learning node embeddings. 
Another recent content-aware approach leverages the attention mechanism of the transformer models over the graph edges and introduces the Heterogeneous Graph Transformer (HGT) \cite{hu2020heterogeneous}.
These works have been vastly explored with heavy connected graphs such as scientific literature graphs.
However, limited research has been done in integrating heterogeneous graph information in the context of scientific news articles.
Our contribution is mainly focused on using GNN techniques to the context of news articles evaluation and assessment. 

\subsection{News Articles Embeddings}
Various methods have been proposed to create meaningful feature vectors for news articles. 
Most of them are focused on Natural Language Processing (NLP) methods for analysing the news. NLP methods are deployed by creating textual representations of the implicit attributes of an article such as its title, its corpus and its images \cite{kowsari2019text, lau2016empirical}.
Other works leverage both the content based attributes as well as contextual information such as article references, social media impact, sources' attributes and related news \cite{zhang2018structured, rappaz2019dynamic}. 
As mentioned, limited research has been made to present and model this problem as a graph problem.
In our contribution, we aim to use contextual information from the network of the cited scientific literature combined with content aware representations, in order to create high performance article embeddings.


\section{Scientific news article network}
\label{sec:network}
The dataset of the scientific news articles that is used for all the experiments is extracted from the NewsTeller platform\footnote{\url{https://newsteller.io}}, which is a research-driven platform to analyze news.
NewsTeller platform manages a large collection of news articles around various topics and domains. 
It enhances each news article with indicators about its content, its social media impact and its referenced sources \cite{romanou2020scilens}.
For this work, we extracted news articles that are related with scientific topics such as: \textit{COVID19, vaccination, healthcare, artificial intelligence, recycling}. 
We filtered these articles by taking the ones with at least one reference to the scientific literature.
The final dataset contains scientific news articles in the English language from various online media platforms along with its cited scientific literature.
The  time  frame  of  the  data  collection,  in  the context of this paper,  covers the 4 month period from 2020-08-01 to  2020-11-31.
The detailed descriptive statistics about the elements of the network are presented in Table \ref{tab:dataset}.

As depicted in Figure \ref{fig:network}(a), the network is comprised of five types of nodes: the scientific news articles, the topics of the scientific news articles, the cited scientific literature, the authors of the cited literature and the institutions of the cited papers, as well as the different relationships between them.
These meta relations between the entities of the used dataset are presented in Figure \ref{fig:network}(b).
Notably, the dataset does not include the citations of the papers to other scientific papers, keeping the utilized network simpler.  
Moreover, since a news article can be assigned to many different news topics, the relation type between article and topic nodes is described as a \textit{many-to-many} relation. 


\begin{table}[t]
  \small
  \caption{Scientific news articles network statistics (above); nodes \& edges statistics (below).}
  \label{tab:dataset}
  \begin{tabular}{cccc}
    \toprule
    Entities & Entity types & Edges & Relation types\\
    \midrule
    5,569 & 5 & 9,547 & 4 \\
  \bottomrule
\end{tabular}

~
\vfill
~

\begin{tabular}{lc}
    \toprule
    Dataset & \#nodes\\
    \midrule
    \#topics & 23 \\
    \#articles & 472 \\
    \#papers & 1,242 \\
    \#authors & 3,464 \\
    \#institutes & 368 \\
  \bottomrule
\end{tabular}
\begin{tabular}{lc}
    \toprule
    Dataset & \#edges\\
    \midrule
    \#cites & 1,421 \\
    \#has\_topic & 1,086 \\
    \#is\_author\_of & 3,576 \\
    \#is\_affiliated\_with & 3,464 \\
    &  \\
  \bottomrule
\end{tabular}
\end{table}

\section{Methodology}
\label{sec:methodology}
In this section we present the methodology followed in order to provide meaningful entity representations for the scientific news articles network.
We define the baseline graph neural network model and propose state-of-the-art approaches inspired by the existing literature.

In order to create graph representations for the presented scientific news network, we implemented a baseline graph neural network for relational graphs (R-GCN) as proposed by \citet{schlichtkrull2018modeling}. 
For the link prediction task, R-GCN is comprised of a graph auto-encoder model.
The encoder creates contextual representations for each entity, and a DistMult \cite{yang2014embedding} decoder produces a score for every potential edge in the graph, using these hidden node representations.
We implemented the R-GCN encoder with a single embedding layer. 
The encoder has been regularized through edge dropout which is applied before normalization, with a dropout rate of 0.4 for all edge types. 
The model uses an Adam optimizer and it is trained using full-batch gradient descent techniques.   

We test this model by comparing it with two content aware methods that leverage the attention mechanism.
The first method is a content aware heterogeneous graph neural network model (HetGNN) as proposed by \citet{zhang2019heterogeneous}.
The produced model is comprised of two modules.
The first module extracts a content embedding for each node using a recurrent neural network on the various attributes related to the node.
The second module utilizes another recurrent neural network, to aggregate these content embeddings for each neighboring node and applies an attention mechanism to measure the impact of heterogeneous node types, creating the final embedding.

The second implemented content aware model is a transformer based model as proposed by \citet{hu2020heterogeneous}. 
This model uses dedicated representations for each different type of nodes and edges and produces node- and edge-type dependent attention mechanism.
Another contribution of this model is that it tackles the temporal nature of the graph by capturing the dynamic structural dependency with arbitrary window sizes.
For this model we use the published date as an additional feature of the news article entities.

Both content aware models use an Adam optimizer and are trained using mini-batch gradient descent.
For fair comparison, we set the embedding size to 128 for all the above approaches.

\begin{table*}[t]
  \small
  \caption{Performance of different models on the two link prediction use cases on the test set.}
  \label{tab:results}
  \begin{tabular}{l|cccc|cccc}
    \toprule
    & \multicolumn{4}{c}{Use case A} & \multicolumn{4}{c}{Use case B} \\
    & MRR & \multicolumn{3}{c}{Hits @} & MRR & \multicolumn{3}{c}{Hits @}\\
    Model &  & 1 & 3 & 10 &  & 1 & 3 & 10 \\
    \midrule
    R-GCN & 0.388 & 0.239 & 0.288 &  0.544  & 0.689 & 0.618 & 0.690 & 0.728 \\
    HetGNN & \textbf{0.412} & \textbf{0.317} & 0.422 & 0.546 & 0.697 & 0.668 & 0.702 & 0.732\\
    HGT & 0.408 & 0.311 & \textbf{0.426} & \textbf{0.589} & \textbf{0.701} & \textbf{0.670} & \textbf{0.713} & \textbf{0.735}\\
  \bottomrule
\end{tabular}
\end{table*}

\section{Experiments \& Results}
\label{sec:results}
We assess the aforementioned models on the downstream task of link prediction for two distinct use cases that are described in Section \ref{sec:intro}.
We split the network edges into a training and a test set with a ratio set to of 5 : 1 for all experiments. 
Both cases were evaluated in the test set with two commonly used metrics: Mean Reciprocal Rank (MRR) and Hits at $n$ (Hits@n). 
In order to meet the original implementations of the models, we report the filtered MRR which is typically considered more reliable and filtered Hits at 1, 3 and 10 positions.

For both content aware models, we used as input features the titles of the scientific news article and the paper.
Thus, for each node we used pre-trained XLNet \cite{yang2019xlnet} to get the representations of each word in its title. 
In continue, we calculate the weighted average of words' attention to get the title representation for each paper and news article, as proposed by \citet{hu2020heterogeneous}.

For the \textbf{Use Case A}, we create entity representations on the scientific news articles network as described in Section \ref{sec:methodology} and we aim to predict the edges connecting the news articles with the scientific literature. 
The results of this experiment are presented in Table \ref{tab:results}. 
Based on the used methodology, we see promising results on both content-aware models; the HetGNN and HGT models, having as node input features the titles of the news articles and the papers.

Similarly, in \textbf{Use Case B}, we also model the scientific news articles network as described in Section \ref{sec:methodology} and we try to predict the edges connecting the news articles with the topic entities. 
As mentioned in the dataset description in Section \ref{sec:network}, each article can be characterised by more than one topics. 
Popular topics in our dataset are the following: \textit{COVID-19, Trump, vaccine, artificial intelligence, healthcare, space} etc. 
It is clear from the semantics of the topics that many \textit{COVID-19} related articles can be also associated with the topic of \textit{vaccine}. 
As presented in Table \ref{tab:results}, the content aware methods display increasing performance with the transformer model achieving better scores in all metrics.

\section{Conclusion}
\label{sec:conclusion}
In this work we showcase different methods for modeling heterogeneous graphs for the case of the scientific news article network. 
We implement different approaches that leverage both the structural nature of the graph as well as the specific attributes for each entity type. 
We show that there is significant potential for further research and improvement on news misinformation and assessment, harnessing the power of graph neural networks and attention mechanisms.
Furthermore, we show that modeling of news related networks can provide numerous benefits around news assessment and fact-checking.
The findings and performance of the implemented models, show promising applications and research regarding the mining and modeling of large heterogeneous graphs. 

Future work could be performed on applying content aware models on more diverse news graphs and implementing more complex attributes for each entity and relation type.
Such attributes can be the paper abstract and the corpus of the news article.
Numerous variations of these experiments could also be developed as future research by incorporating additional network meta relations.
These meta relations could be the citations of news articles to each other or any other type of connection between the news articles and the entities that they quote or mention.
Moreover, further improvements can be made by applying node disambiguation on the cited literature author nodes, a topic that was not addressed in this work.

\balance

\bibliographystyle{ACM-Reference-Format}
\bibliography{sample-base}

\end{document}